\documentclass[aps,twocolumn,groupedaddress,showpacs]{revtex4}
\usepackage{graphicx}
\begin{document}

\title{Optical absorption spectra in ${\rm SrCu_{2}O_{3}}$ two-leg spin ladder }

\author{Nobuyasu Haga and Sei-ichiro Suga}
\affiliation{Department of Applied Physics, Osaka University, Suita, Osaka 565-0871, Japan}
\date{\today}
\begin{abstract}
We calculate the phonon-assisted optical-absorption spectra in ${\rm SrCu_{2}O_{3}}$ two-leg spin-ladder systems. 
The results for two models proposed for ${\rm SrCu_{2}O_{3}}$ are compared.  In the model including the effects of a cyclic four-spin interaction, the shoulder structure appears at $\sim 978 {\rm cm^{-1}}$ and the peak appears at $\sim 1975 {\rm cm^{-1}}$ in the spectrum for polarization of the electric field parallel to the legs. In the other model which describes a pure two-leg ladder, the peak appears around the lower edge of the spectrum at $\sim 1344 {\rm cm^{-1}}$. 
The feature can be effective in determining the proper model for $\rm SrCu_{2}O_{3}$. 
\end{abstract}
\maketitle
%
Intensive studies for $S=1/2$ two-leg spin-ladder systems have revealed fascinating aspects of elementary excitations as well as thermodynamic properties of the system \cite{dagotto}. 
Using several theoretical methods, it was shown that $S=0$ and $S=1$ two-triplet bound states lie below the two-triplet continuum in addition to the one-triplet mode \cite{koto1,sach,koto2,jb,moni1,moni2,uhrig,SCES,nun}.  
Furthermore, the effects of a cyclic four-spin interaction on the $S=1/2$ two-leg spin-ladder systems have attracted a great amount of attention recently \cite{mizuno1,sh,cyc,ins,mizuno2,hh,hn,hmh,lst,mvm,Haga2}. 
From the analysis of the experimental results for ${\rm La_{6}Ca_{8}Cu_{24}O_{41}}$ observed by inelastic neutron-scattering experiments, it was corroborated that a cyclic four-spin interaction is necessary to explain the observed dispersion relation of  the one-triplet mode \cite{ins}.

For ${\rm SrCu_2O_3}$, which is considered to be a typical $S=1/2$ two-leg spin-ladder material, two models were proposed to reproduce the temperature dependence of the susceptibility.   
One model includes the effects of a cyclic four-spin interaction \cite{mizuno1}, while the other model describes a pure two-leg ladder \cite{Johnson}. 
It seems, thus, difficult to decide the proper minimal model only from thermodynamic quantities such as the susceptibility. 
On the basis of the observations, dynamical structure factors have been investigated for the two models \cite{mizuno2, Haga2}. Although the dispersion relation for the model including the effects of a cyclic four-spin interaction becomes flatter, the difference in the distribution of the weights between the two models may be difficult to be observed by inelastic neutron-scattering experiments \cite{Haga2}. 
Other methods to detect characteristics of dynamical properties are required to decide the proper model for ${\rm SrCu_2O_3}$.

The midinfrared optical-absorption spectra measured, {\it e.g.}, by optical conductivity in low-dimensional cuprates can be successfully analyzed in terms of the phonon-assisted absorption mechanism [23-31, 9, 10]. In this mechanism, an absorbed photon simultaneously creates a ${\rm Cu-O}$ bond-stretching phonon and a multitriplet excitation with a singlet coupling. This simultaneous excitation enables us to probe a phonon-weighted sum of the $S=0$ multitriplet excitation from the entire Brillouin zone. 
In fact, it was confirmed from the optical-conductivity measurements that the $S=0$ two-triplet bound state exists below the two-triplet continuum in a $S=1/2$ two-leg spin-ladder material ${\rm (Ca, La)_{14}Cu_{24}O_{41}}$ [31, 9, 10].

In this paper, we investigate the optical-absorption spectra caused by the phonon-assisted two-triplet excitation in ${\rm SrCu_{2}O_{3}}$ two-leg spin-ladder systems on the basis of the proposed two minimal models. 
 Using a continued fraction method based on the Lanczos algorithm \cite{GB87}, we first calculate the wave-number-resolved $S=0$ two-triplet excitation spectra . We then calculate the optical-absorption spectra and discuss characteristics of the results for the two models of ${\rm SrCu_{2}O_{3}}$.

We start our discussion with the following Hamiltonian: 
\begin{eqnarray}
{\cal H}
&=& J_{\|}\sum_{i=1}^{N/2} \left( 
      \mbox{\boldmath$S$}_{1,i} \cdot \mbox{\boldmath$S$}_{1,i+1}
  + \mbox{\boldmath$S$}_{2,i} \cdot \mbox{\boldmath$S$}_{2,i+1} \right)
                                                            \nonumber \\
&+& J_{\bot}\sum_{i=1}^{N/2} 
      \mbox{\boldmath$S$}_{1,i} \cdot \mbox{\boldmath$S$}_{2,i} \nonumber \\
&+& J_{\rm cyc}\sum_{i=1}^{N/2} \{ 4[
(\mbox{\boldmath$S$}_{1,i} \cdot \mbox{\boldmath$S$}_{1,i+1})
(\mbox{\boldmath$S$}_{2,i+1} \cdot \mbox{\boldmath$S$}_{2,i}) \nonumber \\
&+& (\mbox{\boldmath$S$}_{1,i} \cdot \mbox{\boldmath$S$}_{2,i}) 
(\mbox{\boldmath$S$}_{1,i+1} \cdot \mbox{\boldmath$S$}_{2,i+1})  \nonumber \\
&-& (\mbox{\boldmath$S$}_{1,i} \cdot \mbox{\boldmath$S$}_{2,i+1})
(\mbox{\boldmath$S$}_{1,i+1} \cdot \mbox{\boldmath$S$}_{2,i})] \nonumber \\
&+& (\mbox{\boldmath$S$}_{1,i} \cdot \mbox{\boldmath$S$}_{1,i+1}) + 
(\mbox{\boldmath$S$}_{1,i+1} \cdot \mbox{\boldmath$S$}_{2,i+1}) + 
(\mbox{\boldmath$S$}_{2,i+1} \cdot \mbox{\boldmath$S$}_{2,i}) \nonumber \\
&+& (\mbox{\boldmath$S$}_{1,i} \cdot \mbox{\boldmath$S$}_{2,i+1}) +
(\mbox{\boldmath$S$}_{1,i+1}\cdot\mbox{\boldmath$S$}_{2,i}) + 
(\mbox{\boldmath$S$}_{2,i} \cdot \mbox{\boldmath$S$}_{1,i}) 
\}                                                            \nonumber \\
&+& J_{\rm diag} \sum_{i=1}^{N/2} 
     [(\mbox{\boldmath$S$}_{1,i} \cdot \mbox{\boldmath$S$}_{2,i+1})+
    (\mbox{\boldmath$S$}_{1,i+1}\cdot\mbox{\boldmath$S$}_{2,i}) ], 
\label{ham1} 
\end{eqnarray}
%
\noindent
where $\mbox{\boldmath$S$}_{l,i}$ denotes the $S=1/2$ spin operator in the $i$th rung of the $l=1, 2$ chain, $N$ is the total number of spins, and $J_{\|}$ and $J_{\bot}$ are the coupling constants along the leg and rung, respectively. 
The coupling constants for a cyclic four-spin interaction and a diagonal interaction are denoted as $J_{\rm cyc}$ and $J_{\rm diag}$, respectively. 
The coupling constants used in the two models for ${\rm SrCu_2O_3}$ are (i) $J_{\bot}=150 {\rm meV}, J_{\|}=195 {\rm meV}, J_{\rm cyc}=18 {\rm meV}, J_{\rm diag}=3 {\rm meV}$ (Ref. 11) and (ii) $J_{\bot}=86 {\rm meV}, J_{\|}=172 {\rm meV}, J_{\rm cyc}=J_{\rm diag}=0$ (Ref. 22). 
The periodic boundary condition is applied along the chain.

The wave-number-resolved $S=0$ two-triplet excitation spectra for polarization of the electric field parallel to the legs ($\mbox{\boldmath$E$} \, \| \, {\rm legs}$) can be expressed as 
%
\begin{eqnarray}
R^{\rm leg}(q_x,q_y;\omega) 
&=& -\frac{1}{\pi} \Im \langle \Psi_{0}| {\cal A^{\rm leg}}_{q_x,q_y}^{\dagger}   \frac{1}{z-{\cal H}} {\cal A^{\rm leg}}_{q_x,q_y} 
  |\Psi_{0} \rangle  \nonumber \\
&=& F^{\rm leg}(q_x,q_y) C^{\rm leg}(q_x,q_y;\omega)  , 
\label{r1} 
\end{eqnarray}
%
\noindent
where $q_x$ and $q_y(=0, \pi)$ are the wave numbers along the leg and rung, respectively, $|\Psi_{0}\rangle$ is the eigenfunction of the lowest eigenvalue $E_0$, and $z=\omega+i\varepsilon+E_{0}$. 
The operator ${\cal A}_{q_x,q_y}^{\rm leg} = (1/\sqrt{N}) \sum_{l,j}e^{{\rm i}(q_x j + q_y l)} \mbox{\boldmath$S$}_{l,j}\cdot\mbox{\boldmath$S$}_{l,j+1}$ is the Fourier transform of the operator for the locally exciting $S=0$ two-triplet state on the neighboring sites in a chain. 
We set $\hbar =1$ and $\varepsilon = 3.0 \times 10^{-2}$. 
The energy is measured in units of $J_{\|}$. 
In the expression (2), $C^{\rm leg}(q_x,q_y;\omega)$ can be represented in the form of the continued fraction, which can be calculated numerically by a Lanczos algorithm \cite{GB87}. 
The total contribution of $C^{\rm leg}(q_x,q_y;\omega)$ for fixed $q_x$ and $q_y$ is normalized to unity, because the following sum rule has to be satisfied: 
$F^{\rm leg}(q_x,q_y)=\int^{\infty}_{0} {\rm d}\omega R^{\rm leg}(q_x,q_y;\omega)$. 
We use the two sets of coupling constants corresponding to the two models for ${\rm SrCu_2O_3}$: (i) $J_{\bot}/J_{\|} = 0.769$, $J_{\rm cyc}/J_{\|} = 0.0923$, and $J_{\rm diag}/J_{\|} = 0.0153$ (Ref. 11) and (ii) $J_{\bot}/J_{\|} = 0.5$ and $J_{\rm cyc} = J_{\rm diag} = 0$ (Ref. 22).

In Fig. 1, $R^{\rm leg}(q_x,0; \omega)$ and $R^{\rm leg}(q_x,\pi;\omega)$ in the two models are shown for $N=28$. The weight is proportional to the area of the full circle. 
In model (i) the convergences have the relative errors of about $O(10^{-10})$ for $\omega < 3$ and $O(10^{-2}) - O(10^{-5})$ for $\omega > 4$, while in model (ii) they have the relative errors of about $O(10^{-12})$ for $\omega < 3$ and $O(10^{-3}) - O(10^{-5})$ for $\omega > 4$. 
In both models, the large weights for given $q_x$ lie in the lowest excited states  in $q_x \geq 0.4\pi$. 

\begin{figure}[htbp]
\begin{center}
\includegraphics[trim=0 0.6cm 0 0,clip,width=0.48\textwidth]{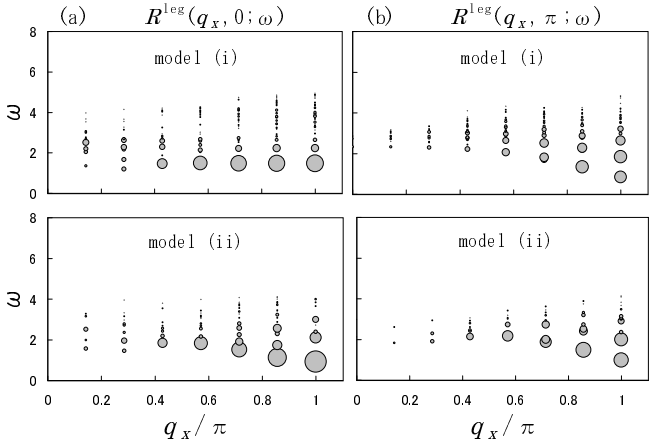}
\caption{
(a) $R^{\rm leg}(q_x,0;\omega)$ and (b) $R^{\rm leg}(q_x,\pi;\omega)$ in model (i) and model (ii) for $N=28$.
The weight is proportional to the area of the full circle. }
\end{center}
\end{figure}

To discuss whether such lowest excited states form a lower edge of the excitation continuum or an isolated mode of the bound state, we next investigate the finite-size effects of the poles and their residues \cite{com} of the continued fraction $C^{\rm leg}(q_x,q_y;\omega)$ \cite{taka,ys}. As reported in Refs. 34-36 and 18, a pole which belongs to an excitation continuum tends to have an appreciable size dependence of at least either its position or its residue. 
Since the size dependence of the position of the pole is hardly seen, we only show the size dependence of the residue in the following.   
In Fig. 2, we first show $R^{\rm leg}(q_x,0;\omega)$ and the size dependence of the lowest excited states of $C^{\rm leg}(q_x,0;\omega)$ for $J_{\bot}/J_{\|} = 1$ and $J_{\rm cyc} = J_{\rm diag} = 0$. In $q_x>0.5\pi$, the size dependence is scarcely seen. Thus, the lowest excited states in $q_x>0.5\pi$ form an isolated mode of the $S=0$ two-triplet bound state, while those in  $q_x<0.5\pi$ form the lower edge of the excitation continuum. The results are consistent with those obtained for the other two methods \cite{uhrig,nun}. 

\begin{figure}[htbp]
\begin{center}
\includegraphics[trim=0 0.6cm 0 0,clip,width=0.48\textwidth]{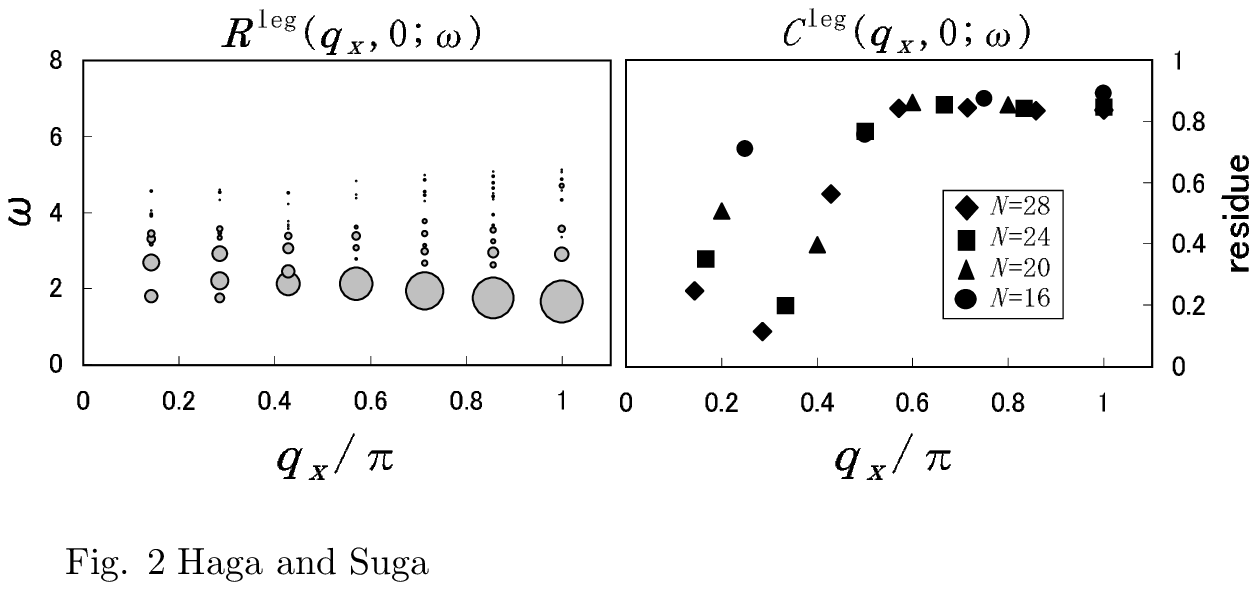}
\caption{
$R^{\rm leg}(q_x,0;\omega)$ for $N=28$ and the size dependence of the residues of the lowest excited states for $C^{\rm leg}(q_x,0;\omega)$ in $J_{\bot}/J_{\|} = 1$ and $J_{\rm cyc} = J_{\rm diag} = 0$.
}
\end{center}
\end{figure}

In Fig. 3, we show the size dependence of the residue for the lowest excited states. In $C^{\rm leg}(q_x,\pi;\omega)$, the residues for both models have a noticeable size dependence in $0 \leq q_x \leq \pi$, indicating that the lowest excited states of the $q_y=\pi$ mode become the lower edges of the excitation continuum in the thermodynamic limit. 
In $C^{\rm leg}(q_x,0;\omega)$ for model (ii), on the other hand, the size dependence in $q_x>0.5\pi$ is very weak, while the residue in $q_x<0.5\pi$ decreases with increasing $N$. Therefore, the lowest excited states of the $q_y=0$ mode in model (ii) form an isolated mode in $q_x>0.5\pi$, and those in $q_x<0.5\pi$ become the lower edge of the continuum. 
In $C^{\rm leg}(q_x,0;\omega)$ for model (i), the residues around $q_x \sim 0.6\pi$ and $0.8\pi$ decrease slightly with increasing $N$. 
Although it is difficult to draw a definite conclusion, the observations may indicate that the lowest excited states of the $q_y=0$ mode in model (i) become the lower edge of the excitation continuum in $q_x>0.6\pi$. 
In fact, the size dependence of the residue in $q_x>0.6\pi$ is quite similar to that for the $S=1/2$ isotropic Heisenberg chain in $q>0.6\pi$ \cite{ys}, where the lowest excited states form the lower edge of the two-spinon continuum. 
Note that in Ref. 10, it is argued that the $S=0$ two-triplet bound state exists in a larger $J_{\rm cyc}$. 
To develop more definite finite-size analysis, a larger system has to be calculated. 

\begin{figure}[htbp]
\begin{center}
\includegraphics[trim=0 0.6cm 0 0,clip,width=0.48\textwidth]{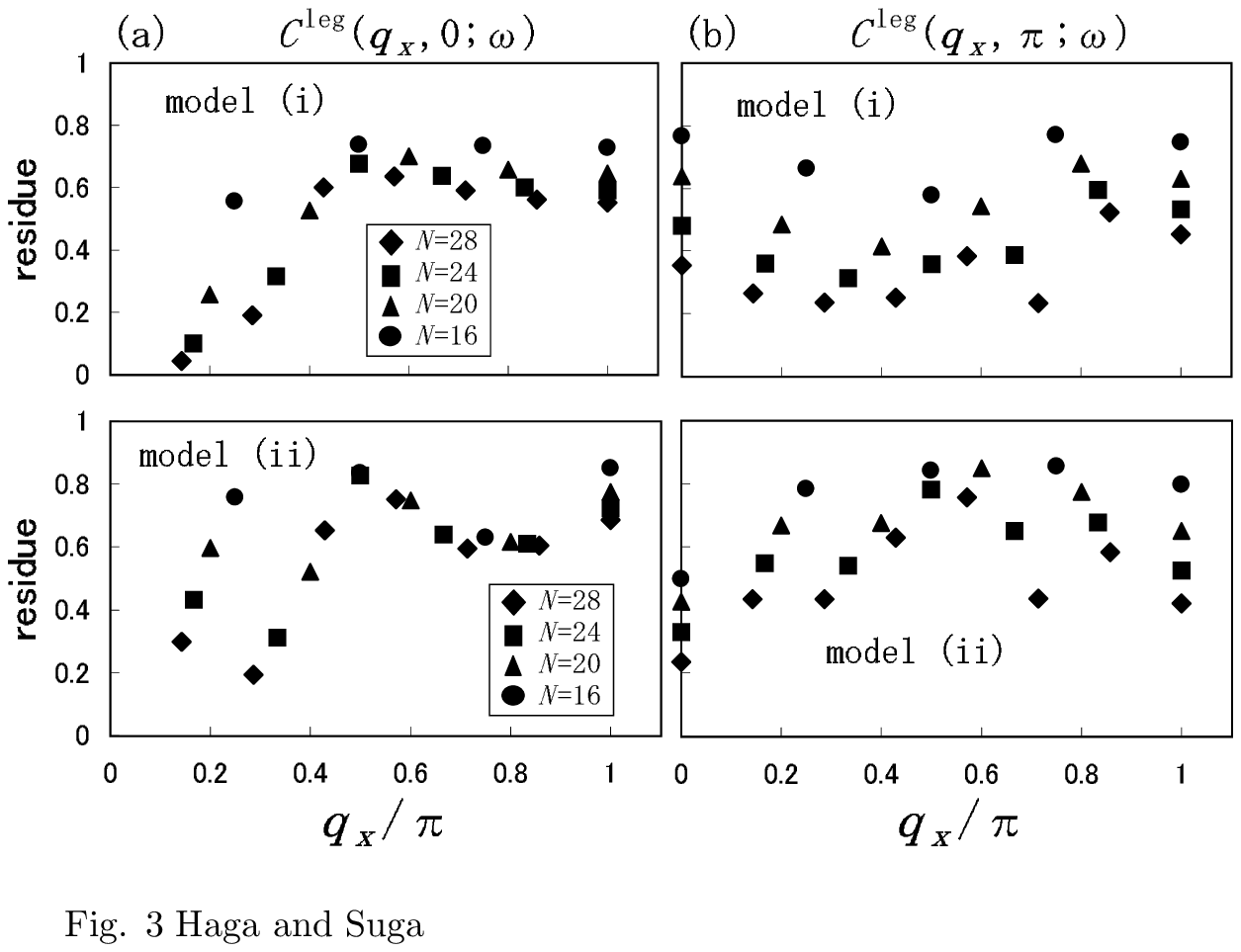}
\caption{
The finite-size effects of the residues of the lowest excited states for (a) $C^{\rm leg}(q_x,0;\omega)$ and (b) $C^{\rm leg}(q_x,\pi;\omega)$ in model (i) and model (ii).
}
\end{center}
\end{figure}

The optical-absorption coefficient for $\mbox{\boldmath$E$} \, \| \, {\rm legs}$ is given by \cite{lz,le,gel,les}
%
\begin{equation}
\alpha^{{\rm leg}}(\omega) = \alpha_{\rm o} \omega I^{{\rm leg}}(\omega-\omega_{\rm o}), 
\label{sp} 
\end{equation}
%
where $\alpha_{\rm o}$ is a constant depending on the material, $\omega_{\rm o}$ is the frequency of the ${\rm Cu-O}$ bond-stretching phonon, and 
$I^{\rm leg}(\omega) = \sum_{q_x,q_y} f_{q_x,q_y}^{\rm leg} R^{\rm leg}(q_x,q_y;\omega)$ with the phonon form factors $f_{q_x,0}^{\rm leg} = \sin^4(q_x/2)$ and $f_{q_x,\pi}^{\rm leg} = \sin^2(q_x/2)+\sin^4(q_x/2)$. 
To calculate $I^{{\rm leg}}(\omega-\omega_{\rm o})$, we apply the spline interpolation to the lowest excited states of $R^{\rm leg}(q_x,q_y;\omega)$ in the  $q_x$ direction.

In Fig. 4, we show $I^{\rm leg}(\omega-\omega_{\rm o})$ thus obtained for the two models in $N=28$. 
The dashed line and dotted line represent the spectra from the $q_y=0$ mode and  $q_y=\pi$ mode, respectively, and the solid line indicates their sum.  A noticeable difference between the results for the two models appears. 
In $I^{\rm leg}(\omega-\omega_{\rm o})$ of model (i), the shoulder appears at $\omega-\omega_{\rm o} \sim 0.7$ and the peak appears at $\omega-\omega_{\rm o} \sim 1.2$. In $I^{\rm leg}(\omega-\omega_{\rm o})$ of model (ii), on the other hand, the peak appears at the lower edge of the spectrum: $\omega-\omega_{\rm o} \sim 1.0$. 
In model (i), the dispersion relation for the lowest excited states of the $q_y=0$ mode becomes flatter, which yields the peak at $\omega-\omega_{\rm o} \sim 1.2$. The shoulder structure is caused by the lowest excited states of the $q_y=\pi$ mode, which lies below those of the $q_y=0$ mode in $q_x \geq 0.8\pi$. 
Note that when we set $J_{\rm cyc}=0$ in model (i), the lowest excited states of the $q_y=\pi$ mode lie above those of the $q_y=0$ mode and no shoulder structure emerges. Therefore, such shoulder and peak structures in model (i) are caused by the cyclic four-spin interaction. 
In model (ii), the dispersion relation of the lowest excited states in the $q_y=0$ mode lies below that of the $q_y=\pi$ mode. Thus, in model (ii) no shoulder appears and the peak structure is seen at the lower edge of the spectrum. 
\begin{figure}[htbp]
\begin{center}
\includegraphics[trim=0 1cm 0 0,clip,width=0.4\textwidth]{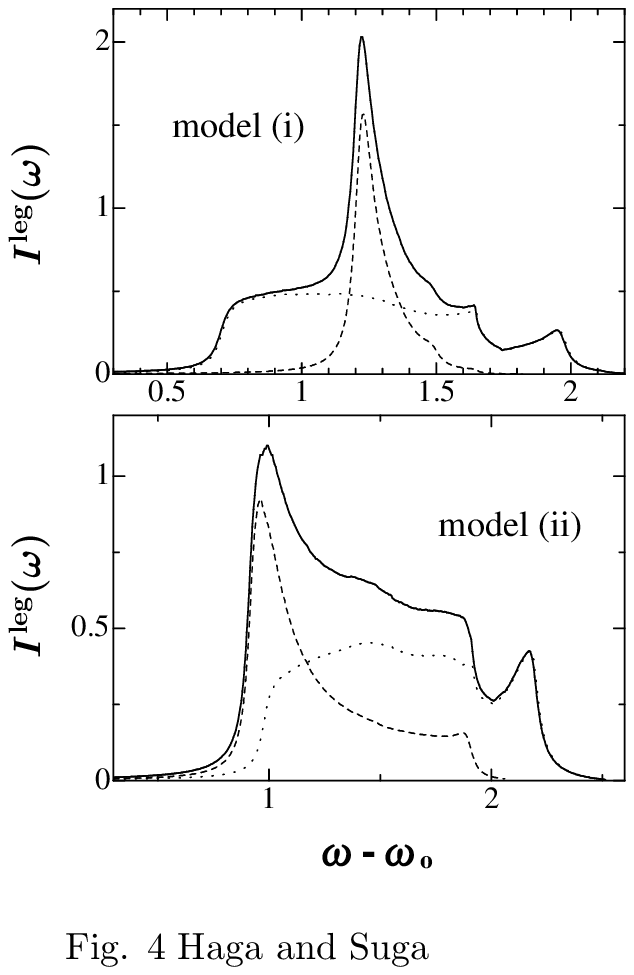}
\caption{
The optical-absorption spectra for $\mbox{\boldmath$E$} \, \| \, {\rm legs}$ in  model (i) and model (ii) for $N=28$. The dashed lines and dotted lines represent the spectra from the $q_y=0$ mode and $q_y=\pi$ mode, respectively, and the solid lines indicate their sum. }
\end{center}
\end{figure}

In Fig. 5(a) we show the size dependence of the excitation energies for the {\it shoulder} and peak, and in Fig. 5(b) we show the size dependence of their weights. As mentioned above, the {\it shoulder} is caused by the $q_y=\pi$ mode (see the dotted lines in Fig. 4) and the peak is caused by the $q_y=0$ mode (see the dashed lines in Fig. 4). 
In model (i), the extrapolated excitation energies for the {\it shoulder} and peak at $N \rightarrow \infty$ are $\omega-\omega_{\rm o} \sim 0.240$ and $\omega-\omega_{\rm o} \sim 0.874$, respectively. 
The extrapolated weights of the {\it shoulder} and peak at $N \rightarrow \infty$ are $\sim 0.0996$ and $\sim 0.474$, respectively. 
Therefore, in model (i) the shoulder and peak structures in the optical-absorption spectrum are intrinsic. 
In model (ii), the extrapolated excitation energies for the {\it shoulder} and peak at $N \rightarrow \infty$ almost coincide and take the values $\omega-\omega_{\rm o} \sim 0.540$ and $0.536$ with corresponding weights of $\sim 0.0553$ and $0.876$. 
Thus, in model (ii) only one peak appears around the lower edge in the optical-absorption spectrum. 

\begin{figure}[htbp]
\begin{center}
\includegraphics[trim=0 0.6cm 0 0,clip,width=0.48\textwidth]{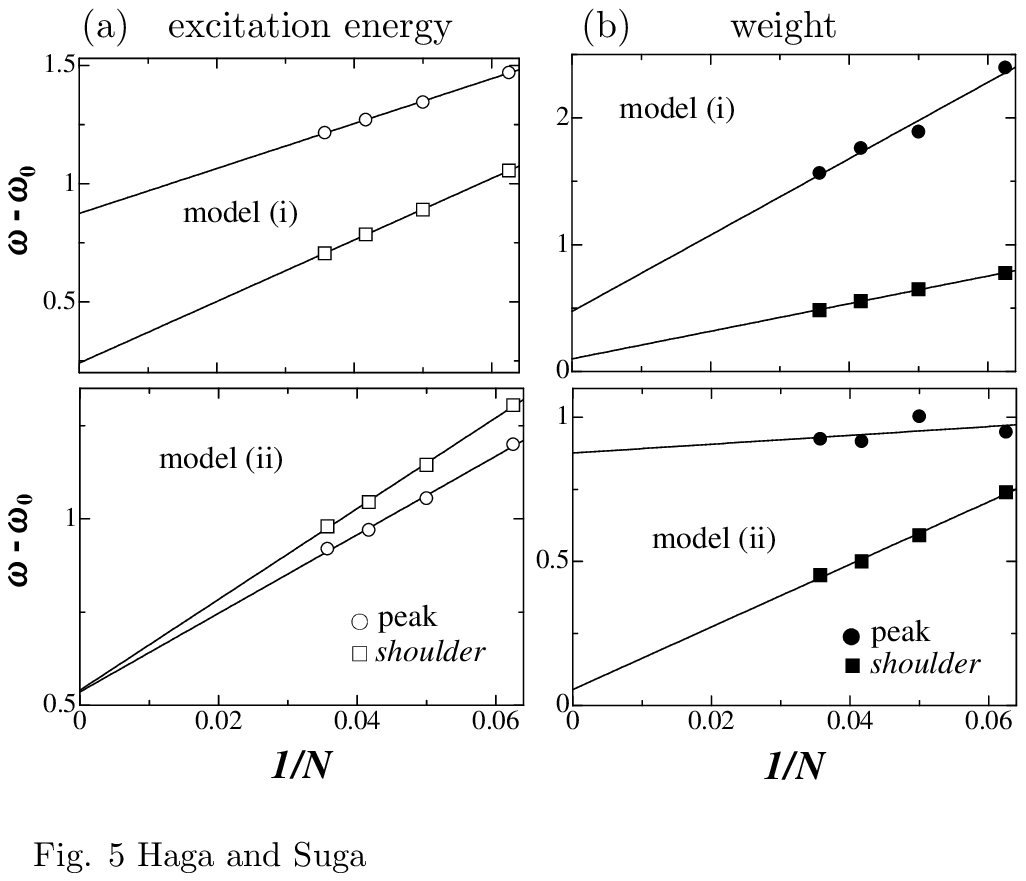}
\caption{
The size dependence of (a) the excitation energies for the peaks (see the dashed lines in Fig. 4) and {\it shoulders} (see the dotted lines in Fig. 4), and of (b) their weights.  
}
\end{center}
\end{figure}

Using $J_{\|}=195 {\rm meV}$ in model (i) \cite{mizuno1}, $J_{\|}=172 {\rm meV}$ in model (ii) \cite{Johnson}, and $\omega_{\rm o}=600{\rm cm^{-1}}$ commonly evaluated in cuprates \cite{legex}, the excitation energies for the shoulder and peak in the model (i) can be evaluated as $978 {\rm cm^{-1}}$ and $1975 {\rm cm^{-1}}$, respectively, while the peak in the model (ii) can be evaluated as $1344 {\rm cm^{-1}}$. 
Since $\alpha^{\rm leg}(\omega)$ can be observed experimentally by the optical conductivity, such different characteristics between the optical-absorption spectra in the two models are probably detectable. 
The observed findings can be effective in determining the appropriate minimal model for $\rm SrCu_{2}O_{3}$.

In summary, we have investigated the phonon-assisted optical-absorption spectra on the basis of the two models proposed for ${\rm SrCu_{2}O_{3}}$. 
Characteristics of the lowest excited states have been discussed in connection with the shoulder and peak structures in the optical-absorption spectra. 
In the model including the effects of a cyclic four-spin interaction, the shoulder and peak structures appear at $\sim 978 {\rm cm^{-1}}$ and $\sim 1975 {\rm cm^{-1}}$, respectively, in the optical-absorption spectrum for $\mbox{\boldmath$E$} \, \| \, {\rm legs}$. 
In the other model which describes a pure two-leg ladder, no shoulder appears and the peak appears around the lower edge of the optical-absorption spectrum for $\mbox{\boldmath$E$} \, \| \, {\rm legs}$ at $\sim 1344 {\rm cm^{-1}}$.

Our computational programs are based on TITPACK version 2 by H. Nishimori. 
Numerical computations were carried out at the Yukawa Institute Computer Facility, Kyoto University, and the Supercomputer Center, the Institute for Solid State Physics, University of Tokyo. 
This work was supported by a Grant-in-Aid for Scientific Research from the Ministry of Education, Culture, Sports, Science, and Technology, Japan.


\begin{references}
%
\bibitem{dagotto}
E. Dagotto, Rep. Prog. Phys. {\bf 62}, 1525 (1999), and references therein.  
%
\bibitem{koto1}
O. P. Sushkov and V. N. Kotov, Phys. Rev. Lett. {\bf 81}, 1941 (1998).  
%
\bibitem{sach}
K. Damle and S. Sachdev, Phys. Rev. B {\bf 57}, 8307 (1998). 
%
\bibitem{koto2}
V. N. Kotov, O. P. Sushkov and R. Eder, Phys. Rev. B {\bf 59}, 6266 (1999). 
%
\bibitem{jb}
C. Jurecka and W. Brenig, Phys. Rev. B {\bf 61}, 14307 (2000). 
%
\bibitem{moni1}
S. Trebst, H. Monien, C. J. Hamer, Z. Weihong, and R. R. P. Singh, Phys. Rev. Lett. {\bf 85}, 4373 (2000). 
%
\bibitem{moni2}
W. Zheng, C. J. Hamer, R. R. P. Singh, S. Trebst, and H. Monien, Phys. Rev. B {\bf 63}, 144410 (2001). 
%
\bibitem{uhrig}
C. Knetter, K. P. Schmidt, M. Gr\"{u}ninger, and G. S. Uhrig, Phys. Rev. Lett. {\bf 87}, 167204 (2001). 
%
\bibitem{SCES}
M. Gr\"uninger, M. Windt, T. Nunner, C. Knetter, K. P. Schmidt, G. Uhrig, T. Kopp, A. Freimuth, U. Ammerahl, B. B\"uchner, and A. Revcolevschi, J. Phys. Chem. Solids {\bf 63}, 2167 (2002). 
%
\bibitem{nun}
T. S. Nunner, P. Brune, T. Kopp, M. Windt, and M. Gr\"uninger, Phys. Rev. B {\bf 66}, 180404 (2002).  
%
\bibitem{mizuno1}
Y. Mizuno, T. Tohyama, and S. Maekawa, J. Low Temp. Phys.{\bf 117}, 389 (1999).
%
\bibitem{sh}
T. Sakai and Y. Hasegawa, Phys. Rev. B {\bf 60}, 48 (1999). 
%
\bibitem{cyc}
S. Brehmer, H.-J. Mikeska, M. M\"{u}ller, N. Nagaosa, and S. Uchida, Phys. Rev. B {\bf 60}, 329 (1999). 
%
\bibitem{ins}
M. Matsuda, K. Katsumata, R. S. Eccleston, S. Brehmer, and H.-J. Mikeska, Phys. Rev. B {\bf 62}, 8903 (2000). 
%
\bibitem{mizuno2}
Y. Mizuno, T. Tohyama, and S. Maekawa, J. Phys. Chem. Solids {\bf 62}, 273 (2001). 
%
\bibitem{hh}
Y. Honda and T. Horiguchi, cond-mat/0106426. 
%
\bibitem{hn}
K. Hijii and K. Nomura, Phys. Rev. B {\bf 65}, 104413 (2002). 
%
\bibitem{Haga2}
N. Haga and S. Suga, Phys. Rev. B {\bf 66}, 132415 (2002). 
%
\bibitem{mvm}
M. M\"{u}ller, T. Vekua, and H.-J. Mikeska, Phys. Rev. B {\bf 66}, 134423 (2002). 
%
\bibitem{hmh}
T. Hikihara, T. Momoi, and X. Hu, Phys. Rev. Lett. {\bf 90}, 087204 (2003). 
%
\bibitem{lst}
A. L\"{a}uchli, G. Schmid, and M. Troyer, Phys. Rev. B {\bf 67}, 100409(R) (2003). 
%
\bibitem{Johnson}
D. C. Johnston, Phys. Rev. B {\bf 54}, 13009 (1996). 
%
\bibitem{lz}
J. Lorenzana and G. A. Sawatzky, Phys. Rev. Lett. {\bf 74}, 1867 (1995); Phys. Rev. B {\bf 52}, 9576 (1995). 
%
\bibitem{suz}
H. Suzuura, H. Yasuhara, A. Furusaki, N. Nagaosa, and Y. Tokura, Phys. Rev. Lett. {\bf 76}, 2579 (1996). 
%
\bibitem{gru1}
M. Gr\"uninger, J. M\"unzel, A. Gaymann, A. Zibold, H. P. Geserich, and T. Kopp, Europhys. Lett. {\bf 35}, 55 (1996).
%
\bibitem{le}
J. Lorenzana and R. Eder,  Phys. Rev. B {\bf 55}, R3358 (1997). 
%
\bibitem{perkins2}
J. D. Perkins, R. J. Birgeneau, J. M. Graybeal, M. A. Kastner, and D. S. Kleinberg, Phys. Rev. B {\bf 58}, 9390 (1998). 
%
\bibitem{gel}
D. Garcia, J. Eroles, and J. Lorenzana, Phys. Rev. B {\bf 58}, 13574 (1998). 
%
\bibitem{les}
J. Lorenzana, J. Eroles, and S. Sorella, Phys. Rev. Lett. {\bf 83}, 5122 (1999).%
\bibitem{gru2}
M. Gr\"uninger, D. van der Marel, A. Damascelli, A. Erb, T. Nunner, and T. Kopp, Phys. Rev. B {\bf 62}, 12422 (2000).
%
\bibitem{legex}
M. Windt, M. Gr\"uninger, T. Nunner, C. Knetter, K. P. Schmidt, G. S. Uhrig, T. Kopp, A. Freimuth, U. Ammerahl, B. B\"uchner, and A. Revcolevschi, Phys. Rev. Lett. {\bf 87}, 127002 (2001).
%
\bibitem{GB87}
E. R. Gagliano and C. A. Balseiro, Phys. Rev. Lett. {\bf59}, 2999 (1987).
%
\bibitem{com} For finite systems, $C^{\rm leg}(q_x,q_y;\omega)$ consists of a finite number of Lorenzians. The pole and residue depict the position of the Lorenzian and the integrated value of each Lorenzian with respect to $\omega$. 
%
\bibitem{taka}
M. Takahashi, Phys. Rev. B {\bf 50}, 3045 (1994). 
%
\bibitem{ys}
H. Yokoyama and Y. Saiga, J. Phys. Soc. Jpn {\bf 66}, 3617 (1997). 
%
\bibitem{hs1}
N. Haga and S. Suga, Phys. Rev. B {\bf 65}, 014414 (2002). 
%
%
%
\end{references}

\end{document}